\documentclass[fleqn,twoside]{article}
\usepackage{espcrc2}

\usepackage{psfrag}
\usepackage{amsfonts}
\usepackage[dvips,dvipdfm]{graphicx}
\usepackage[figuresright]{rotating}

\def\Journal#1#2#3#4{{#1} #2 (#4) #3}
\def\PRD{{Phys. Rev.} D}

\newcommand{\AmS}{{\protect\the\textfont2
  A\kern-.1667em\lower.5ex\hbox{M}\kern-.125emS}}

\title{Preliminary results in unquenched Numerical Stochastic Perturbation Theory
\vskip-4.6cm\hfill\small DESY-03-147; UPRF-2003-22\vskip4.6cm}

\author{F. Di Renzo\address[PARMA]
					{Dipartimento di Fisica, Universit\`a di Parma and INFN, 
				   Gruppo Collegato di Parma, Italy},
 				A. Mantovi\addressmark[PARMA],	
 				V. Miccio\addressmark[PARMA],
        L. Scorzato\address[Boston]
        	{DESY - Hamburg, Germany}}

\begin{document}

\begin{abstract}
Introducing fermionic loops contributions in Numerical 
Stochastic Perturbation Theory was mainly motivated by the 
proposal to compute $2$-$3$ loops  for renormalization constants 
(and improvement coefficients). This is feasible because the computational 
overhead of unquenching NSPT is by far lower than for non perturbative 
simulations. We report on 
first, preliminary results for the quark propagator (basically the 
third loop for the critical mass) and discuss the status of the computation of other quantities. 
\vspace{1pc}
\end{abstract}

\maketitle

\section*{INTRODUCTION}

Numerical Stochastic Perturbation Theory (NSPT) is a numerical method to perform Lattice Perturbation Theory which has proven to be very powerful in the quenched approximation of Lattice QCD. In recent years \cite{lat00} the unquenched version was introduced, whose first (benchmark) results were encouraging \cite{lat02}. The goal, which appears feasible, is to have NSPT competitive on any $2$-$3$ loops computation of quantities like renormalization constants (and possibly improvement coefficients, where needed) for potentially any fermionic action. An obvious first benchmark one can go through is the quark propagator for Wilson action. In other words, one is computing $Z_{\psi}$ to $3$ loops, for example matching to the RI' scheme. In the following we only report on a byproduct, the critical mass for $N_f=2$.

\section{The computational set-up}
For a comprehensive account of the formulae of unquenched NSPT we refer the reader to \cite{lat00}. We briefly discuss instead the computational set-up. Results are obtained on a $32^4$ lattice with $N_f=2$ Wilson action. The relevant configurations are generated on a APEmille crate ($128$ FPU's for $64$ GFlops of peak performance). Smaller lattices to keep finite size effects under control are demanded to a PC cluster. Stability with respect to single vs double precision has been tested. Gauge is fixed (where needed) to the Landau condition (Fourier acceleration has turned out to be very helpful). Given the replication of the fields at every perturbative order, a single configuration on a $32^4$ lattice at $3$ loops ({\em i.e.} $\beta^{-3}$) level amounts to $1.8$ GB. We are collecting a large amount of confi\-gurations in order to set up a repository out of which any fermionic observable can be computed in perturbation theory to $3$ loops. 

\subsection{Status of unquenched computations}

As already stated, the configurations we have been generating up to now are at $N_f=2$. Given the polynomial dependence of fermionic observa\-ble on $N_f$ in perturbation theory, it will be easy to fit it once configurations will be available also at other value of $N_f$. There is anyway a computation already on its way just now, which is that of the third order of the residual mass counter\-term in Lattice Heavy Quark Effective Theory. The goal is to extend the unquenched result for the b-quark mass \cite{ourHQET} to three loops accuracy for the perturbative subtraction which is needed to keep renormalons effects under control. An $N_f=2$ computation will be enough for this purpouse. In recent years other methods to compute the b-quark mass in completely non-perturbative ways have been proposed \cite{otherHQET}. The only unquenched computations available at the moment are anyway still the ones relying on HQET and perturbative subtractions of the residual mass, so we still think that it is worthwhile to proceed. The residual mass, as it was the case for the quenched case, is being computed out of the computation of the matching between lattice and potential scheme couplings, which in turns relies on computations of big Wilson loops. 
Quark bilinears are the obvious first quantities we are computing (again, at $N_f=2$ level, at the moment) in the quark sector of the theory. As a matter of fact also four fermions operators look feasible. 

\section{The critical mass}
\begin{figure}[t]
%	\vspace{-.3cm}
	\begin{center}
%		\psfrag{Placchetta}[cc][cc]{$\langle\Box\rangle_{\beta^{-1}}$}
%		\psfrag{Reticolo}[cc][cc][.9]{$L$ --- [lattice size]}
%		\psfrag{Titolo1}[bb][bb][.9]{1-loop coefficient for}
%		\psfrag{Titolo2 }[cc][cc][.9]{the plaquette mean-value $\langle\Box\rangle$}
%		\psfrag{Data interpolation}[cc][cc][.9]{Data interpolation}
%		\psfrag{Analytical result}[cc][cc][.9]{Analytical result}
		\includegraphics[scale=0.45]{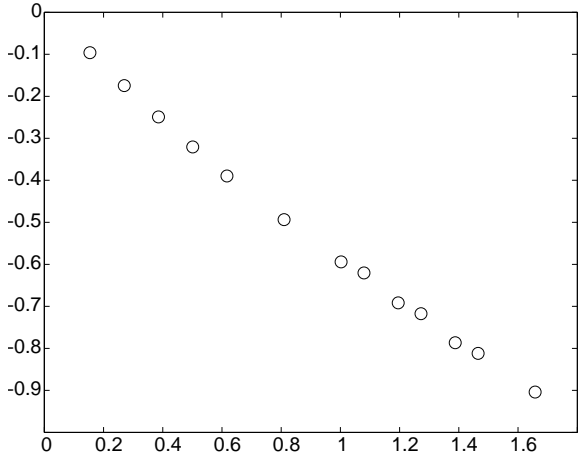}
	\end{center}
	\vspace{-1.1cm}
	\caption{$-\Sigma_0$ at order $\beta^{-1}$ as a function of $p^2$ ($N_f=2$) in renormalized perturbation theory (critical mass counterterm as known from analytical computations plugged in at one and two loops level).}
	\label{fig:o1}
	\vspace{-.3cm}
\end{figure}
\begin{figure}[t]
%	\vspace{-.3cm}
	\begin{center}
%		\psfrag{Placchetta}[cc][cc]{$\langle\Box\rangle_{\beta^{-1}}$}
%		\psfrag{Reticolo}[cc][cc][.9]{$L$ --- [lattice size]}
%		\psfrag{Titolo1}[bb][bb][.9]{1-loop coefficient for}
%		\psfrag{Titolo2 }[cc][cc][.9]{the plaquette mean-value $\langle\Box\rangle$}
%		\psfrag{Data interpolation}[cc][cc][.9]{Data interpolation}
%		\psfrag{Analytical result}[cc][cc][.9]{Analytical result}
		\includegraphics[scale=0.45]{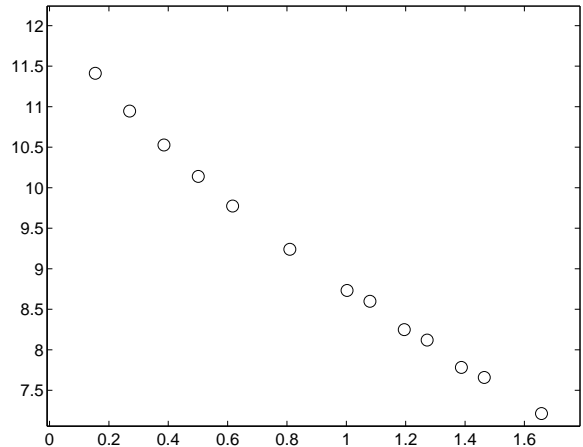}
	\end{center}
	\vspace{-1.1cm}
	\caption{$-\Sigma_0$ at order $\beta^{-3}$ as a function of $p^2$ ($N_f=2$). The result has not yet been extrapolated in the stochastic time-step $\tau \rightarrow 0$.}
	\label{fig:o3}
	\vspace{-.3cm}
\end{figure}

The quark propagator (which we denote by $S(p^2,m)$) is our first benchmark computation among the quark bilinears. In NSPT approach one actually computes $S$ and then proceeds to invert to get
\begin{eqnarray}
	\Gamma_2(p^2,m) & = & S(p^2,m)^{-1} \\ \nonumber
	& = & i \not p - m - \Sigma(p^2,m)
\end{eqnarray}
which is the quantity which actually encodes all the information. $\Sigma$ in turns goes like
\begin{equation}
	\Sigma(p^2,m) = \Sigma_0 + i \not p \Sigma_1(p^2,m) + m \Sigma_2(p^2,m)
\end{equation}
where $\Sigma_0$ is a pure lattice artifact (actually diverging as $a^{-1}$). At $m=0$ it is the well-known critical mass, the additive renormalization one has to subtract from the bare lagrangian mass in order to obtain the bare mass that gets multiplicatively renormalized. It is the only quantity we are actually going to report on here (again, for $N_f=2$). Still it is worthwhile to note the features all $\Sigma_i$ share with respect to our NSPT computation. As already stated, one first computes $S$, then inverts to get $\Gamma$ and then projects on the (Dirac space) identity and the $\gamma_{\mu}$ in order to get the $\Sigma_i$. Of course one takes advantage of the colour simmetry to improve the accuracy. The dependences on $p^2$ and $m$ are completely known at one loop level, while those on $p^2$ and $m=0$ are known from continuum perturbation theory (notice that from these one can elaborate on $p^2$ and $m$ dependence at $2$ and $3$ loops). The $m=0$ results are the ones we are interested in here (they also define the point where one would like to compute in general the renormalization factors $Z$). We used two IR regulators in order to get the $m=0$ results. We both plugged small $m\approx0$ values in and subtracted zero modes in the $m=0$ simulations. Results showed that we can control the procedure to a good accuracy. On the lattice at a finite value of the lattice spacing $\Sigma \neq \Sigma(p^2)$. The dependence on the lattice invariants one can construct as powers of $p a$ are actually the natural handles for the $a\rightarrow 0$ 
limit. \\
The first and second loop contribution to the critical mass are analytically known \cite{m_cr}. For $N_f=2$ they read
\begin{equation}
	- \Sigma_0 = 2.6057 \,\beta^{-1} + 4.293 \,\beta^{-2} + \ldots
\end{equation}
\begin{figure}[t]
%	\vspace{-.3cm}
	\begin{center}
%		\psfrag{Placchetta}[cc][cc]{$\langle\Box\rangle_{\beta^{-1}}$}
%		\psfrag{Reticolo}[cc][cc][.9]{$L$ --- [lattice size]}
%		\psfrag{Titolo1}[bb][bb][.9]{1-loop coefficient for}
%		\psfrag{Titolo2 }[cc][cc][.9]{the plaquette mean-value $\langle\Box\rangle$}
%		\psfrag{Data interpolation}[cc][cc][.9]{Data interpolation}
%		\psfrag{Analytical result}[cc][cc][.9]{Analytical result}
		\includegraphics[scale=0.45]{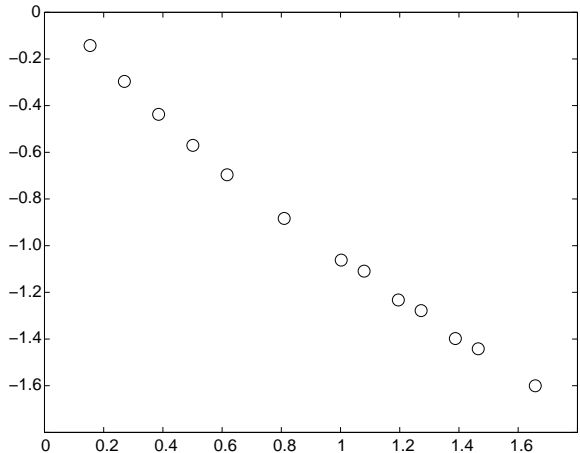}
	\end{center}
	\vspace{-1.1cm}
	\caption{$-\Sigma_0$ at order $\beta^{-2}$ as a function of $p^2$ ($N_f=2$) in renormalized perturbation theory (same framework as in Fig.~\ref{fig:o1}).}
	\label{fig:o2}
	\vspace{-.3cm}
\end{figure}
Notice that once one knows these values, one actually has to plug them in as counterterms in NSPT simulations (to keep their contributions into account at higher loops). In Fig.~\ref{fig:o1} we plot the component along the identity of $\Gamma_2$ ({\em i.e.} $\Sigma_0$, given the $m=0$ value) at order $\beta^{-1}$. This is plotted against $p^2$. The curve is not smooth since (as we have already pointed out) we know this is not a function of $p^2$ alone: the value extrapolated at $p=0$ is the quantity one is interested in. Notice that this is what one would call renormalized (with respect to the additive mass renormalization) perturbation theory: one has plugged the first two counterterms in, so $\Sigma_0$ gets cancelled at one and two loops level. The confirmation of this at two loops level is clear in Fig.~\ref{fig:o2} (the same as Fig.~\ref{fig:o1} at $\beta^{-2}$ order). From Fig.~\ref{fig:o3} one can inspect that the result for the third order is within reach. We are not quoting a number at this preliminary stage, but the order of magnitude of the result is clear. Notice, by the way, that we plot the result from a very reduced set of configurations at a fixed time-step (in NSPT one performs simulations at a fixed value of the time-step used in integrating Langevin equation and then has to extrapolate to the continuum-time stochastic dynamics). We add just a few comments to have a rough idea of the order of magnitude of this new contribution to the (perturbative) critical mass. If one compares the perturbative result for the critical mass at three loops level with the one determined with non-perturbative methods, one finds that results are still quite far away from each other. To have an idea of the quite poor convergence of the series in the bare coupling: at $\beta = 5.5$ the contribution of order $\beta^{-3}$ is some $50$\% that of order $\beta^{-2}$ (which is in turns $30$\% that of order $\beta^{-1}$).

\section*{Conclusions and acknowledgments}
By performing a three loops computation of the Wilson action critical mass at $N_f=2$ we showed that three loops computations are indeed feasible in unquenched NSPT. The goal is now to compute at this order various renormalization constants. \\
We thank G. Burgio for very interesting and helpful discussions.

\begin{thebibliography}{9}

\bibitem{lat00}
F. Di Renzo and L. Scorzato, Nucl. Phys. B Proc. Suppl. 94 (2001) 567.
\bibitem{lat02}
F. Di Renzo, V. Miccio and L. Scorzato, Nucl. Phys. B Proc. Suppl. 119 (2003) 1003.
\bibitem{ourHQET} V. Gimenez, L. Giusti, G. Martinelli and F. Rapuano, JHEP 03 (2000) 018. 
F. Di Renzo and L. Scorzato, JHEP 02 (2001) 020.
\bibitem{otherHQET} R. Sommer, Nucl. Phys. B Proc. Suppl. 119 (2003) 185. G. M. de Divitiis, M. Guagnelli, R. Petronzio, N. Tantalo and F. Palombi, hep-lat/0305018.
\bibitem{m_cr} E. Follana, H. Panagopoulos, \Journal{\PRD}{63}{17501}{2001}; S. Caracciolo, 
A. Pelissetto, A. Rago, \Journal{\PRD}{64}{94506}{2001}.

\end{thebibliography}
\end{document}